\documentclass[pra,twocolumn,a4paper,nofootinbib,superscriptaddress]{revtex4}

\usepackage{hyperref}
\usepackage{graphicx}
\usepackage{amsmath}
\usepackage{amssymb}
\usepackage{color}
\usepackage{amsthm}
\usepackage{amsfonts}
\newcommand{\bra}[1]{{\left\langle{#1}\right\vert}}
\newcommand{\ket}[1]{{\left\vert{#1}\right\rangle}}

\begin{document}

\title{Reducing the overhead for quantum computation when noise is biased}

\author{Paul Webster}
\affiliation{Centre for Engineered Quantum Systems, School of Physics, The University of Sydney, Sydney, NSW 2006, Australia}
\author{Stephen D. Bartlett}
\affiliation{Centre for Engineered Quantum Systems, School of Physics, The University of Sydney, Sydney, NSW 2006, Australia}
\author{David Poulin}
\affiliation{D{\'e}partment de Physique, Universit{\'e} de Sherbrooke, Sherbrooke, Qu{\'e}bec J1K 2R1, Canada}

\date{5 November 2015}

\begin{abstract}
We analyse a model for fault-tolerant quantum computation with low overhead suitable for situations where the noise is biased.  The basis for this scheme is a gadget for the fault-tolerant preparation of magic states that enable universal fault-tolerant quantum computation using only Clifford gates that preserve the noise bias.  We analyse the distillation of $|T\rangle$-type magic states using this gadget at the physical level, followed by concatenation with the 15-qubit quantum Reed-Muller code, and comparing our results with standard constructions.  In the regime where the noise bias (rate of Pauli $Z$ errors relative to other single-qubit errors) is greater than a factor of 10, our scheme has lower overhead across a broad range of relevant noise rates.
\end{abstract}

\maketitle

\section{Introduction}

Fault-tolerant quantum computation provides a means to process quantum information with faulty devices using quantum error-correcting codes together with logical gate constructions that do not propagate errors, in such a way that a quantum computation of arbitrary length can occur provided the error rate is below a threshold~\cite{AGP}.  For two decades now, the focus for research into quantum architectures has been to increase this error threshold as high as possible.  The high-threshold schemes of Knill~\cite{Knill} and Raussendorf, Harrington, and Goyal~\cite{RHG} have error thresholds of around $1\%$, and there has not been significant improvement in these threshold values in the subsequent decade of research.  Unfortunately, these high-threshold schemes possess dauntingly large overheads, that is, a significant excess in the number of physical qubits and gates necessary to construct the fault-tolerant logical operations.  For example, the high-threshold scheme for quantum computation based on surface codes~\cite{RHG} has an overhead that increases with problem size, but typical numbers are of the order of $10^6$ or more physical qubits per logical qubit when magic state distillation is included~\cite{Fowler}.  There is considerable motivation, then, for the design of fault tolerant schemes with low overhead~\cite{Poulin,BravyiHaah,Gottesman,JOConnor,Jones,Meier}.

%Quantum computation holds the promise of efficiently performing tasks that are classically intractable, such as factoring~\cite{Shor}.  While noise will in general destroy the coherence that offers the computational advantage, quantum error correction can be used to lower the effective noise rate on the encoded information~\cite{Shor2}.  Using such codes as part of a quantum circuit with noisy operations requires all operations to be performed on the encoded information in a fault tolerant manner, that is, in a way that prevents errors propagating through the circuit during its correct operation~\cite{Preskill}.  With the right code and a fault-tolerant circuit, a quantum computation of arbitrary size can in principle be performed provided that the noise rate is below a threshold.

In this paper, we explore a method for reducing this overhead in systems where the noise is biased, specifically, where dephasing (the $Z$ error rate) is dominant over other errors.   Our results are motivated by previous studies of architectures with high error threshold in such a noise bias regime~\cite{Pres}.  Although the increase in the error threshold in these noise-biased architectures is relatively modest compared with standard constructions, we demonstrate that the overhead can be reduced by approximately a factor of four across a wide range of noise regimes.  

One of the most significant contributions to the large overhead in existing fault-tolerant quantum computing schemes is the distillation of high-fidelity magic states that enable a universal logic gate set using only Clifford operations, and several investigations have explored how to reduce this particular cost~\cite{BravyiHaah,JOConnor}.  Building on these previous results, we propose and analyse a simple fault-tolerant gadget capable of directly preparing encoded magic states in a system affected by biased noise.  This gadget has significant error correction capabilities for $Z$ errors, but ignores all other errors.   We derive expressions for the logical error rates on the encoded magic state preparation in terms of the circuit parameters, and demonstrate that for bare error rates of less than $10^{-3}$ and biases of at least $10^2$, the logical error rate of the encoded magic state is lower than the bare error rate.  
%While our gadget is nondeterministic for encoded $|T\rangle$-type magic states, in the instances where it fails it prepares a different encoded magic state on the equator of the Bloch sphere.  
The simplicity of this gadget should result in reduced overheads when used in a larger fault-tolerant construction. 

To quantify these gains, we consider the overhead involved in using these magic states to prepare high-fidelity magic states in a quantum Reed-Muller code, and compare with standard distillation methods.  Here, we find that for biases of at least a factor of 10, the overhead for $|T\rangle$-type magic state distillation using the scheme at the physical level followed by concatenated layers of the 15-qubit quantum Reed-Muller distillation is almost always lower than using the quantum Reed-Muller distillation directly, for a broad range of physical and target error rates.  Our comparison uses only non-Clifford gate count, and so ignores the fact that our gadget offers a reduction in Clifford gate overhead as well.

The paper is structured as follows.  We introduce the noise model, quantum codes, and gate set for biased noise in Section~\ref{sec:codes}.  In Section~\ref{sec:magicprep}, we present the circuit for fault-tolerant production of magic states, and detail its operation.  Section~\ref{sec:errors} presents an analysis of how errors can arise and propagate through the circuit, including a calculation of the logical noise rate affecting the output encoded state. Section~\ref{sec:distill} provides an illustration of how this gadget could be used within a larger fault-tolerant scheme with lower overhead, by considering the specific case of encoding the output state into a Reed-Muller code for further distillation using standard techniques.  Section~\ref{sec:conclusion} presents some concluding remarks.

\section{Quantum codes and logic gates for biased noise}
\label{sec:codes}

\subsection{Errors and bias}

A standard noise model for studying quantum error correcting codes and fault-tolerant circuits is for single-qubit $X$, $Y$, and $Z$ errors to occur with equal probability, i.e., a uniform single-qubit depolarization channel, independently on all qubits.  While this noise model has a theoretical simplicity, it is not representative of the observed noise on many physical manifestations of a qubit.  In particular, for qubits defined by nondegenerate energy levels with Hamiltonian proportional to $Z$, the noise model is generically described by a dephasing ($Z$-error) rate that is distinct from the rates for relaxation and other non-energy-preserving errors.  Examples include trapped ions~\cite{Nigg}, superconducting qubits~\cite{AliferisSQ}, and electron spins in semiconductors~\cite{Shulman}.  For these qubits, the observed $Z$-error rate can be substantially larger than all other error rates.  

We consider a phenomenological noise model that includes both independent single qubit errors occurring at any location in the circuit (preparations, gates, measurements, and waiting times) as well as correlated two-qubit errors occurring at two-qubit gates.  Unlike in standard noise models, our single qubit errors are described by independent Pauli $Z$ and $X$ errors occurring at different rates.  Let $p_z$ and $p_x$ denote the probability of $Z$ and $X$ errors, respectively, resulting from the noise on a gate, or during a state preparation or measurement.  We assume $p_x < p_z$, that is, that $X$ errors occur with a lower frequency than $Z$ errors, and define the \emph{bias} to be $\eta = p_z/p_x$.  As an example, for a qubit encoded in a pair of electron spins in semiconductor quantum dots~\cite{Shulman}, dephasing dominates over relaxation processes resulting in a noise bias of at least $\eta = 10^3$.  

For the correlated errors, because we will make use of an entangling gate that is diagonal in the $Z$ basis, assume that only correlated $Z$ errors occur on both qubits in a two qubit gate with rate $p_{zz}$.  (Note that correlated $Z$ errors are distinct error processes to independent $Z$ errors occurring on two qubits, which occur with probability $p_z^2 \ll p_z$.)  Capacitive coupling of electron spins in semiconductor quantum dots~\cite{Shulman} provides an example of this situation as well.  For simplicity, we assume that $p_{zz}$ is also much less than $p_z$ and of the same order of magnitude as $p_x$.  

\subsection{Quantum codes for biased noise}

With a noise bias, it is natural to select a quantum error correcting code that offers better properties (specifically, distance) for $Z$ errors than other errors.  The simplest such code is an $n$-qubit repetition code for $Z$ errors, which has distance $n$ for $Z$ errors but no correction capability for other errors.  (That is, it can detect up to $n-1$ $Z$ errors, and correct up to $(n-1)/2$ $Z$ errors, but cannot detect nor correct $X$ errors.) This stabilizer code has $n-1$ stabiliser generators, of the form $X_j X_{j+1}$, for $1 \leq j < n$, and hence encodes one logical qubit. As logical operators we then have $X_L = X_1$, $Z_L = Z^{\otimes n}$.   Code states for the repetition code expressed in the $X$ basis are 
\begin{equation}
	|{+}\rangle_L = |{+}\rangle^{\otimes N} \,, \quad |{-}\rangle_L = |{-}\rangle^{\otimes N} \,.
	\label{eq:RepCode}
\end{equation}

\subsection{Quantum gates for a fault-tolerant construction}

Along with selecting a set of quantum codes, a scheme for fault-tolerant quantum computation requires a universal gate set and a method for performing logical operations in a fault-tolerant way.  To take advantage of the noise bias, we require that all gates commute with $Z$ and thus maintain the direction of the noise bias, i.e., they do not map $Z$ errors into $X$ errors.  

We use a modification of the gate set of Ref.~\cite{Pres}.  The elementary operations that we perform on physical qubits consist of preparation and measurement in the $X$ basis, as well as a two-qubit entangling operation $CZ(\theta)_{ij} = \exp(i \frac{\theta}{2} Z_i\otimes Z_j)$ acting on qubits $i$ and $j$, generated by the Hamiltonian $H_{Z_iZ_j} \propto Z_i \otimes Z_j$.  For $\theta=\pi/2$, this gate is equivalent to the CPHASE gate up to local bias-preserving transformations, where CPHASE is the two qubit Clifford gate with matrix representation $\text{CPHASE} = {\rm diag}(1,1,1,-1)$. (Specifically, CPHASE${}_{ij} = \exp(i\frac{\pi}{4})\exp(-i\frac{\pi}{4}Z_i)\exp(-i\frac{\pi}{4}Z_j)CZ(\theta)_{ij}$).  However, for general rotations $\theta\neq \pi/2$, $CZ(\theta)$ is not a Clifford gate.

This choice of gates is motivated by the noise bias. Specifically, $X$ errors act trivially on preparations and measurements in the $X$ basis, and hence only $Z$ errors affect these operations. Moreover, $CZ(\theta)$ commutes with $Z$, meaning that $Z$ errors on the input will remain $Z$ errors on the output, preserving the noise bias.

With these elementary operations, we can use the results of Ref.~\cite{Pres} to perform fault-tolerant encoded versions of the Clifford operations in the set
\begin{equation}
  \bigl\{P_{|+\rangle_L}, M_{X_L}, M_{Z_L}, \text{CNOT} \bigr\}	
\end{equation}
where $P_{|{+}\rangle_L}$ denote preparation of an encoded qubit in the logical state $|{+}\rangle_L$, $M_{X_L}$ and $M_{Z_L}$ denotes measurement of the logical operators $X_L$ and $Z_L$, respectively, and $\text{CNOT}$ is the logical controlled-NOT operation.  

To obtain a universal gate set, we supplement these Clifford operations with preparation of magic states $P_{|{+i}\rangle_L}$ and $P_{|T\rangle_L}$ through the use of a gadget based on $CZ(\theta)$ gates with $\theta \neq \pi/2$.  Here, $|{+i}\rangle = |0\rangle + i |1\rangle$ and $|T\rangle = |0\rangle + e^{i\pi/4} |1\rangle$ (we omit normalisation for clarity).  The $|T\rangle$ state in particular is not a stabilizer state.  One method to prepare such magic states in a fault-tolerant way is to prepare encoded versions of these states in an appropriate quantum error correcting code, and to use the error correction properties of the code to yield high-fidelity encoded magic states despite having noisy preparations at the physical level.  In the next section, we describe a scheme to implement the encoded magic state preparations $P_{|{+i}\rangle_L}$ and $P_{|T\rangle_L}$ when the noise is biased, in a simple way, using the $CZ(\theta)$ gate and the repetition code.

%Transversal application of a gate within a quantum error correcting code is that such that logical operators may be performed between a pair of logical qubits simply by applying an operator separately between pairs of corresponding bare qubits in each logical qubit. This is of particular interest for fault-tolerant computation, since it is an extremely simple way to prevent propogation of errors through the circuit. However, there does not exist a non-trivial error correcting code for which a universal set of gates for quantum computation may be performed transversally \cite{EastinKnill}. Hence, a set of transversal gates must always be complemented by at least one non-transversal gate.

%A significant set of gates is the Clifford group, defined as the normaliser of the Pauli group. This set is not universal, and may be applied transversally in a large class of quantum codes \cite{JOConnor}. One way to achieve universal quantum computation using these gates is through the injection of additional states, known as ``magic states'' \cite{BraKit}. 

\section{Encoded magic state preparation}
\label{sec:magicprep}

In this section, we present the scheme to prepare encoded magic states such as $|{+i}\rangle_L$ and $|{T}\rangle_L$ in an $n$-qubit repetition code.  
%This state can be injected into a Clifford circuit to perform the $T$ gate (also known as the $\pi/8$ gate), as outlined in Ref.~\cite{Pres}. 

%In general, the circuit uses an $n$ qubit repetition code, and outputs states of the form
%$$|\bar{T}\rangle = |\bar{0}\rangle + e^{i \frac{\pi}{4}} |\bar{1}\rangle = (1+e^{i \frac{\pi}{4}})|\bar{+}\rangle + (1-e^{i \frac{\pi}{4}})|\bar{-}\rangle. $$ Note that, here, and throughout this section, we omit normalisation constants for clarity, as is common in this field.

\begin{figure}
\centering
	\includegraphics[width=1\linewidth]{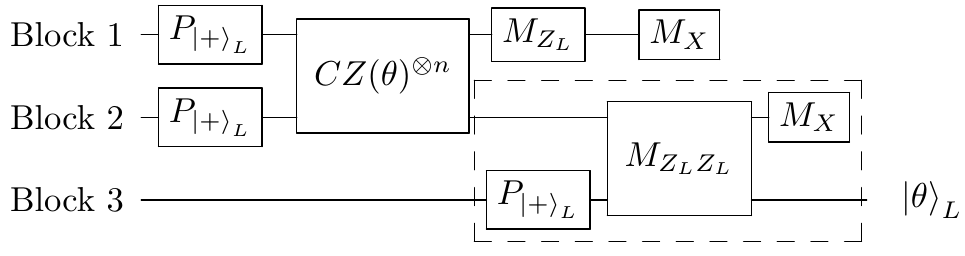}
%\centerline{
%\Qcircuit @C=0.5em @R=.7em {
%\lstick{\mbox{Block 1}} & \gate{P_{\ket{{+}}_L}} & \qw & \multigate{1}{CZ(\theta)^{\otimes n}} & \gate{M_{Z_L}} & \gate{M_{X}} \\
%\lstick{\mbox{Block 2}}& \gate{P_{\ket{{+}}_L}} & \qw & \ghost{CZ(\theta)^{\otimes n}} & \qw & \multigate{1}{M_{Z_L Z_L}} & \gate{M_{X}} \\
%\lstick{\mbox{Block 3}} & \qw  & \qw  & \qw  &  \gate{P_{\ket{{+}}_L}} & \ghost{M_{Z_L Z_L}} & \qw & \qw & \rstick{\ket{\theta}_L}  \gategroup{2}{5}{3}{7}{.7 em}{--} \\
%&\\
%}}
\caption{Representation of the magic state preparation gadget in terms of logical qubits and operators. Each line represents a logical qubit, encoded in an $n$ qubit repetition code. The section marked with a dashed box is used for error correction.}
\label{fig:circuit}
\end{figure}

The fault-tolerant gadget is shown in Fig.~\ref{fig:circuit}.  The basic non-Clifford operation in this gadget is a gate $CZ(\theta)^{\otimes n}$, which entangles blocks 1 and 2 by applying $CZ(\theta)$ as if it were a transversal operation.  Note that $CZ(\theta)$ does \emph{not} act transversally on the repetition code, and so this gate does not preserve the codespace.  However, mapping out of the codespace can be fixed by performing quantum error correction with this code.  
%Specifically, a measurement of $Z_L$ on block 1 collapses the state of block 2 into an encoded magic state, up to errors that are identified by the subsequent $X$ measurement on each of the qubits in block 1.  
The final stage (within the dashed box in Fig.~\ref{fig:circuit}) is an error-correction gadget introduced in Ref.~\cite{Pres}.  It is an adaptation of a one-bit teleportation circuit that teleports the state from block 2 onto block 3 while correcting $Z$ errors.  The output of this final stage is a valid codestate under ideal operations, which depends on the choice of $\theta$ and the intermediate measurement outcomes.  We show that the output for particular values of $\theta$ can be converted using Pauli operations conditioned on these intermediate outcomes to the Clifford magic state $|{+i}\rangle_L$ deterministically, and to the non-Clifford magic state $|{T}\rangle_L$ with some probability (i.e., for certain outcomes).

%We now outline the detailed operation of the gadget.  For clarity, we present the calculations with $n=3$ as the size of the repetition code, but provide results for arbitrary $n$.

\subsection{Magic state preparation}

Blocks $1$ and $2$ are each prepared as $|{+}\rangle_L = |{+}\rangle^{\otimes n}$, and then entangled using the $CZ(\theta)^{\otimes n}$ performed pairwise between the blocks.  The action of this operation is as follows.  On each individual pair of qubits, one from each block, $CZ(\theta)$ acts as
\begin{align}
CZ(\theta) |+\rangle |+\rangle &= e^{-i \frac{\theta}{2} ZZ} (|0\rangle + |1\rangle)(|0\rangle + |1\rangle) \nonumber \\
  &= (II + XX)|0\rangle |\theta\rangle\,,
\end{align}
where we have defined the state $|\theta\rangle = e^{-i\theta/2}|0\rangle + e^{i\theta/2}|1\rangle$.
Therefore,
\begin{equation}
CZ(\theta)^{\otimes n} |{+}\rangle_L|{+}\rangle_L =  [(II +XX) |0\rangle |\theta\rangle]^{\otimes n}\,.
\end{equation}
The $CZ(\theta)^{\otimes n}$ operation does not preserve the codespace of the two blocks.  As we will show, the action of $CZ(\theta)^{\otimes n}$ can be viewed as providing a rotation on the logical state of block 2 conditional on the logical state of block 1, with the addition of systematic correlated $X$ ``errors'' on both blocks as well as correctable $Z$ errors that map out of the codespace.  Both of these effects can be corrected by the gadget.

\begin{figure}
\centering
	\includegraphics[width=0.6\linewidth]{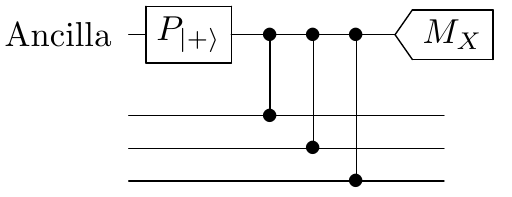}
%\centerline{
%\Qcircuit @C=0.5em @R=.7em {
%\lstick{\mbox{Ancilla}} &  \gate{P_\ket{{+}}} & \qw & \ctrl{2} & \qw & \ctrl{3} & \qw & \ctrl{4} & \qw &  \measuretab{M_X} \\
%& \\
%& \qw & \qw & \control \qw & \qw & \qw & \qw & \qw & \qw & \qw \\ 
%& \qw & \qw & \qw & \qw & \control \qw & \qw & \qw & \qw & \qw \\
%& \qw & \qw & \qw & \qw & \qw & \qw & \control \qw & \qw & \qw \\
%}
%}
\caption{The $M_{Z_L}$ gadget. Each controlled gate is a CPHASE gate acting between a qubit in block $1$ and an ancilla qubit.}
  \label{fig:MZ}
\end{figure}

The measurement of $Z_L$ on the first block will reveal the parity of the number of such correlated $X$ errors, thus collapsing the state on one of two possibilities
\begin{align}
\ket{\Psi_+} &= \sum_{a\in \{0,1\}^n \atop |a| \ {\rm even}} \bigotimes_{i=1}^n (XX)^{a_i}\ket 0  \ket{\theta}  \\
\ket{\Psi_-} &= \sum_{a\in \{0,1\}^n \atop |a| \ {\rm odd}} \bigotimes_{i=1}^n (XX)^{a_i}\ket 0  \ket{\theta}.
\end{align}
The measurement of $Z_L$ on the first block can be performed as in Fig.~\ref{fig:MZ}, and can be repeated $r_z$ times.  (We visit the issue of optimal $r_z$ in the subsequent error analysis in Secs.~\ref{sec:errors} and~\ref{sec:distill}.)

Next, every qubit of the first block is measured in the $X$ basis. Denoting $x_i = \pm 1$ the result of the measurement on the $i$th qubit, the states on block 2 are transformed to
\begin{align}
\ket{\Psi'_+} &= \sum_{a\in \{0,1\}^n \atop |a| \ {\rm even}} \bigotimes_{i=1}^n (x_iX)^{a_i}\ket{\theta} \equiv A_+ \ket\theta^{\otimes n}  \\
\ket{\Psi'_-} &= \sum_{a\in \{0,1\}^n \atop |a| \ {\rm odd}} \bigotimes_{i=1}^n (x_iX)^{a_i}\ket{\theta}\equiv A_- \ket\theta^{\otimes n} ,
\end{align}
where $A_\pm$ implicitly depends on the measurement outcomes $x_i$. Note that $Z_L A_\pm = \pm A_\pm Z_L$ due to the parity constraint. Thus, we can denote the state on block 2 at this stage by $A_\pm \ket\theta^{\otimes n} = A_\pm (e^{i\theta Z/2} \ket +)^{\otimes n}$, and we see that we have used the $CZ(\theta)$ gate to implement a $Z$-rotation, up to known errors.  

\subsection{Error correction}

\begin{figure}
\centering
	\includegraphics[width=0.75\linewidth]{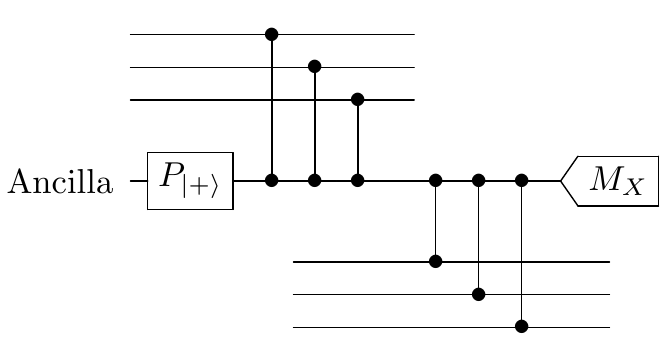}
%\centerline{
%\Qcircuit @C=0.5em @R=.7em {
%& \qw & \qw & \control \qw & \qw & \qw & \qw & \qw & \qw & \qw & \qw \\ 
%& \qw & \qw & \qw & \qw & \control \qw & \qw & \qw & \qw & \qw & \qw \\
%& \qw & \qw & \qw & \qw & \qw & \qw & \control \qw & \qw & \qw & \qw \\
%& \\
%\lstick{\mbox{Ancilla}} &   \gate{P_\ket{{+}}} & \qw & \ctrl{-4} & \qw & \ctrl{-3} & \qw & \ctrl{-2} & \qw & \qw & \qw & \ctrl{2} & \qw & \ctrl{3} & \qw & \ctrl{4} & \qw &  \measuretab{M_X}  \\
%& \\
%&   &   &   &   & \qw & \qw & \qw & \qw
%& \qw & \qw & \control \qw & \qw & \qw & \qw & \qw & \qw & \qw \\ 
%&   &   &   &   & \qw & \qw & \qw & \qw
%& \qw & \qw & \qw & \qw & \control \qw & \qw & \qw & \qw & \qw \\
%&   &   &   &   & \qw & \qw & \qw & \qw
%& \qw & \qw & \qw & \qw & \qw & \qw & \control \qw & \qw & \qw \\
%}
%}
\caption{The $M_{Z_L Z_L}$ gadget. Each controlled gate is a conditional phase gate between data qubits and the ancilla.}
  \label{fig:MZZ}
\end{figure}

The remaining part of the gadget (within the dashed box in Fig.~\ref{fig:circuit}) is an error correction gadget, as introduced in Ref.~\cite{Pres}.  In this final stage, a third register in the state $\ket +_L$ is appended and the system and a measurement of $Z_LZ_L$ is performed on blocks 2 and 3, as shown in Fig.~\ref{fig:MZZ}. Because the operators $A_\pm$ have well-defined commutation/anti-commutation relations with $Z_L$, we can ignore their presence to study the effect of the measurement.  (The effect of these operators is only to flip the sign of the measurement outcome in a deterministic way.)  Moreover, the $Z_LZ_L$ measurement commutes with $(e^{i\theta Z/2})^{\otimes n}$, so we can also ignore the presence of this rotation to study the effect of the measurement.  With these two simplifications, the $Z_LZ_L$ measurement is performed on the state $\ket+_L\ket+_L$.  Denoting the outcome as $(-1)^b$, the result is the preparation of one of two encoded Bell states $\ket0_L\ket 0_L + \ket 1_L\ket 1_L$ for the $b=0$ outcome and $\ket0_L\ket 1_L + \ket 1_L\ket 0_L = (I_LX_L) (\ket0_L\ket 0_L + \ket 1_L\ket 1_L)$ for the $b=1$ outcome. The full state of blocks 2 and 3 at this stage can be written as
\begin{align}
&A_\pm I_L (e^{i\theta Z/2}\otimes I)^{\otimes n} (I_LX_L)^b (\ket0_L\ket 0_L + \ket 1_L\ket 1_L) \nonumber \\
& = A_\pm I_L (e^{i\theta Z/2}\otimes I)^{\otimes n} (I_LX_L)^b (\ket+_L\ket +_L + \ket -_L\ket -_L) \nonumber \\
& = A_\pm I_L (e^{i\theta Z/2}\otimes I)^{\otimes n} (I_LX_L)^b (\ket+^{\otimes 2n}+ \ket -^{\otimes 2n}). 
\end{align}

Finally, each qubit of the second block is measured in the $X$ basis.  For such measurements, using the following identity:
\begin{equation}
	\langle \alpha|\langle \beta| (II + XX) = \begin{cases} 2 \langle \alpha|\langle \beta| & \text{if}\ \alpha = \beta \\ 0 & \text{otherwise} \end{cases}
\end{equation}
for $\alpha,\beta \in \{ {+}, {-} \}$, it is straightforward to show from the definition of the $A_\pm$ operators that, under ideal operation, the $X$ basis measurements on blocks 1 and 2 are either perfectly correlated or perfectly anticorrelated.  Note that if the $X$ basis measurements on block 1 and 2 are not perfectly correlated/anticorrelated in this way, we reject the state.  

To determine the final state on block 3, we use another simple identity:  
\begin{equation}
  \bra\alpha e^{i\theta Z/2} \ket\beta = e^{i\theta/2} + \alpha\beta e^{-i\theta/2}\,, \quad \alpha, \beta \in \{+,-\} \,.
\end{equation}
Denoting $\alpha_i$ the outcome of the $i$th $X$ measurement, and $|\alpha|$ the total number of $+1$ outcomes, the final state is
%\begin{multline}
% \prod_i (e^{i\theta/2} + \beta_i(-1)^b e^{-i\theta/2}) \ket+_L \\ + \prod_i (e^{i\theta/2} - \beta_i(-1)^b e^{-i\theta/2}) \ket -_L
%\end{multline}
\begin{multline}
	\bigl(1+(-1)^b e^{-i\theta}\bigr)^{|\alpha|}\bigl(1-(-1)^b e^{-i\theta}\bigr)^{n-|\alpha|}\ket+_L \\
	+ \bigl(1-(-1)^b e^{-i\theta}\bigr)^{|\alpha|} \bigl(1+(-1)^b e^{-i\theta}\bigr)^{n-|\alpha|}\ket-_L \,.
\end{multline}

The above expression gives a code state depending on the choice of $\theta$ and the intermediate measurement results on block 2.  Certain magic states can be prepared for specific choices of $\theta$, such that the output is Pauli-correctable to a fixed state independent of the intermediate measurement outcomes.  For preparing $|{+i}\rangle_L$, we choose $\theta=\pi/2$, and note that all possible output states are equivalent up to correctable Pauli errors.  

For $|{T}\rangle_L$, choosing $\theta=\pi/4$, the output state is Pauli correctable to the desired state if and only if $n-|\alpha| = |\alpha| \pm 1$.  The number of vectors satisfying this requirement in the field $\mathbb{Z}_2^n$, for odd $n$, is 
\begin{equation}
  \binom{n}{\frac{n-1}{2}} + \binom{n}{\frac{n+1}{2}}= 2 \binom{n}{\frac{n-1}{2}} \,.	
\end{equation}
Thus, the probability of the state being acceptable is:
\begin{equation}
  p_{\rm accept} = 2^{-n} \times 2 \binom{n}{\frac{n-1}{2}} = 2^{1-n} \binom{n}{\frac{n-1}{2}} .
\end{equation}
For $n=3$, this expression gives $p_{\rm accept} = 3/4$; for $n=9$ we have $p_{\rm accept} = 1/2$.  Note that, for $n=3$, the failure channel when all three $X$ measurement outcomes are equal results in the Clifford state $|{+i}\rangle_L$.

%We emphases that other measurement outcomes also provide encoded magic states, which may be useful despite not being Pauli-correctable to a fixed state.  While we discard such outcomes, these states nonetheless provide a potential resource.  For example, with $n=3$ and $\theta=\pi/4$, the instances with 2 similar and 1 different $X$ measurement outcomes yields a $|T\rangle_L$ state, whereas the case with all 3 $X$ measurement outcomes the same produces a magic state $|0\rangle_L + e^{i\pi/8}|1\rangle_L$.  \sdb{Paul:  can you double check?}

\section{Error analysis}
\label{sec:errors}

In this section, we consider how physical errors at the various points in the gadget of Fig.~\ref{fig:circuit} lead to logical $X_L$ and $Z_L$ errors, and place upper bounds on these logical error rates.

\subsection{Logical $X_L$ error bound}

A logical $X_L$ error occurs if an $X$ error affects any of the qubits in the output block.  Working backwards through the steps in the gadget, we see that an $X$ error on block 3 occurs in only two ways:  either an $X$ error occurs on one of the qubits in block 3 (with total probability $\varepsilon_{x,3}$), or the measurement $M_{Z_L Z_L}$ used in the error correction gadget is faulty (with probability $\varepsilon_{x, M_{Z_L Z_L}}$).  Therefore the logical $X_L$ error probability $E_{X_L}$ is upper bounded by
\begin{equation}
  E_{X_L} \leq \varepsilon_{x,3} + \varepsilon_{x, M_{ZZ}}\,.
\end{equation}
To determine $\varepsilon_{x,3}$ on block 3, we note that $X$ errors do not affect the state preparation $P_{\ket{+}_L}$, but a non-trivial $X$ error can occur during any of the $r_{zz}$ CPHASE gates, on any of the $n$ qubits. Therefore,
\begin{equation}
  \varepsilon_{x,3} \leq r_{zz} n p_x\,.
\end{equation}
A faulty $M_{Z_L Z_L}$ measurement can result from a $Z$ error on the majority of $r_{zz}$ ancillas (each of which could occur during any of $2n$ CPHASE gates, or during preparation or measurement), or from an $X$ error on one of the qubits in block 2 (with total probability $\varepsilon_{x,2}$). Thus
\begin{equation}
  \varepsilon_{x, M_{Z_L Z_L}} \leq \binom{r_{zz}}{\frac{r_{zz} + 1}{2}} \big((2n+2)  p_z\big)^{\frac{r_{zz}+1}{2}} + \varepsilon_{x,2}\,.
\end{equation}
An $X$ error on block 2 could occur from a direct $X$ error any of the $n$ qubits of the block, on any of the $n$ $CZ(\theta)$ operations, or on any of $r_{zz}$ CPHASE gates. Additionally, an $X$ error can result from a faulty measurement of $M_{Z_L}$ on block 1, with probability $\varepsilon_{x, M_{Z_L}}$. Thus,
\begin{equation}
  \varepsilon_{x,2} \leq n(r_{zz}+1)p_x + \varepsilon_{x, M_{Z_L}}\,.
\end{equation}
A faulty $M_{Z_L}$ measurement could result from an $X$ error on any of the $n$ $CZ(\theta)$ operations or on any of the $r_z$ CPHASE gates.  Additionally, it could result from a $Z$ error on the majority of $r_z$ ancillas, each of which could occur on any of $n$ CPHASE gates, or during preparation or measurement. Thus,
\begin{equation}
  \varepsilon_{x, M_{Z_L}} \leq n(r_z +1)p_x +  \binom{r_{z}}{\frac{r_{z} + 1}{2}} \bigl((n+2)  p_z\bigr)^{\frac{r_{z}+1}{2}} \,.
\end{equation}

Combining these expressions, as well as choosing the number of repetitions for $M_{Z_L}$ and $M_{Z_L Z_L}$ to be the same, $r_z = r_{zz}\equiv r$, we obtain the following bound on the logical $X_L$ error rate:
\begin{equation}
\label{eq:EXL}
E_{X_L} \leq n(3r+2)p_x + \binom{r}{m} \bigl[\bigl(2(n+1)\bigr)^m + (n+2)^m\bigr]p_z^m \,,
\end{equation}
where we have defined $m=(r+1)/2$.  Note the effect of repeated measurement $r$ is to exponentially suppress the effect of $Z$ errors at the expense of linearly increasing the effect of $X$ errors.

From this expression, we can identify the most significant contribution to the logical $X_L$ error rate $E_{X_L}$ in various noise regimes and choices of $r$.  Choosing $r=n$ (as is standard in related constructions) will maximally suppress the contribution from $Z$ errors, and so the $X$ errors will contribute most significantly to the logical $X_L$ error rate $E_{X_L}$ unless the bias is extremely high ($\gg (25 p_z)^{-1}$).  Choosing $r=1$, both $Z$ and $X$ errors contribute directly (with rate proportional to $p_x$ and $p_z$, respectively) to logical $X_L$ errors.  Therefore, in a noised-biased regime, the optimal choice of $r$ is nontrivial, and we may benefit from choosing $r<n$.  We return to the best choice of $r$ in the next section.  

\subsection{Logical $Z_L$ error bound}

Logical $Z_L$ errors occur if the error correction gadget fails.  The first such failure mode is when a $Z$ error occurs on all $n$ qubits in either blocks 1 or 2; the result is a logical error that cannot be detected by the error correction gadget.  On block 1, $Z$ errors can occur during preparation, measurement, the $CZ(\theta)$ gates, or any of $r_z$ CPHASE gates. The same holds for block 2, but with $r_{zz}$ CPHASE gates instead of $r_z$. Additionally, a single $X$ error on an ancilla for the measurements  $M_{Z_L}$ and $M_{Z_L Z_L}$ could cause $Z_L$ errors on block 1 and block 2, respectively. Such an error can happen on each qubit in any of $r_z$ repetitions $n$ CPHASE gates for  $M_{Z_L}$ or any of $r_{zz}$ repetitions of $2n$ CPHASE gates for  $M_{Z_L Z_L}$. The probability $\varepsilon_{z,1}$ of any of the above possibilities occurring is upper bounded by
\begin{equation}
  \varepsilon_{z, 1} \leq \Bigl(\bigl((r_{z} + 3)+(r_{zz}+3)\bigr)p_z\Bigr)^n + n(r_z + 2r_{zz})p_x \,.
\end{equation}

A second source of logical $Z_L$ errors is the possibility of preparing the incorrect magic state $|\theta\rangle_L$ as a result of a faulty $X$ measurement.  (Note that such failures are distinct from cases where the $X$ measurement results are correct, but correspond to a different angle than the one desired; in such cases the output can be discarded.)   We simplify the calculation by treating any such incorrect angle as a logical $Z_L$ error.  Because the $X$ measurements on block 2 are compared with the $X$ measurements on block 1, all such fault channels will be detected unless correlated $Z$ errors occur on both qubits in a pair on blocks 1 and 2.  When such a pair of errors occurs on blocks 1 and 2, the resulting state is different from the desired $|\theta\rangle_L$ state.   Such a process could happen either as a correlated error in the $CZ(\theta)^{\otimes n}$ gate or as a pair of independent $Z$ errors on a pair of qubits.  (Note that this is the sole error process where the repetition code does not directly protect against $Z$ errors.)   While correlated errors can occur only at a $CZ(\theta)$ gate, independent $Z$ errors on pairs of qubits can occur during preparation, any of the operations, or at measurement.  This yields an upper bound of
\begin{equation}
  \varepsilon_{z,2} \leq n p_{zz}+ n\bigl((r_z +3) p_z\bigr)^2  \,.
\end{equation}
%We note that this error is governed primarily by the first term in the sum, $k=1$, and so the leading order contribution occurs with probability proportional to $p_z^2$.  

Thus, again making the assumption that $r_z = r_{zz} \equiv r$, we obtain a bound on the total logical $Z_L$ error rate for the circuit:
\begin{equation}
\label{eq:EZL}
E_{Z_L} \leq \bigl(2(r+3)p_z\bigr)^n + n p_{zz} + 3n r p_x + n\bigl((r +3) p_z\bigr)^2 \,.
\end{equation}

As with the logical $X_L$ error rate, the effect of repeated measurement $r$ is to exponentially suppress the effect of $Z$ errors in the first term at the expense of increasing the effect of $X$ errors linearly with $r$.  However, here the contribution from $\varepsilon_{z,3}$ (correlated errors that lead to incorrect $X$ measurement results) is not suppressed further by increasing $r$ and will always have a leading order contribution $\sim p_z^2$ regardless of the number of repetitions $r$ or the size of the code $n$.     

The logical $X_Z$ and $Z_L$ error rate expressions motivate some deeper consideration into the best choice of the number of measurement repetitions $r$.  As noted above, the standard approach is to choose $r=n$, which ensures the measurements are fault-tolerant to $Z$ errors on the ancilla and thereby reduce (primarily) the logical $X_L$ error rate.  Depending on the noise bias, however, it may be beneficial to choose $r<n$ and allow an increase in the logical $X_L$ error rate with the aim of reducing the logical $Z_L$ error rate and the associated overhead.  We return to this issue using a specific example in the next section.

\subsection{Results for logical error rates}

\begin{figure}
	\centering
	\includegraphics[width=0.95\linewidth]{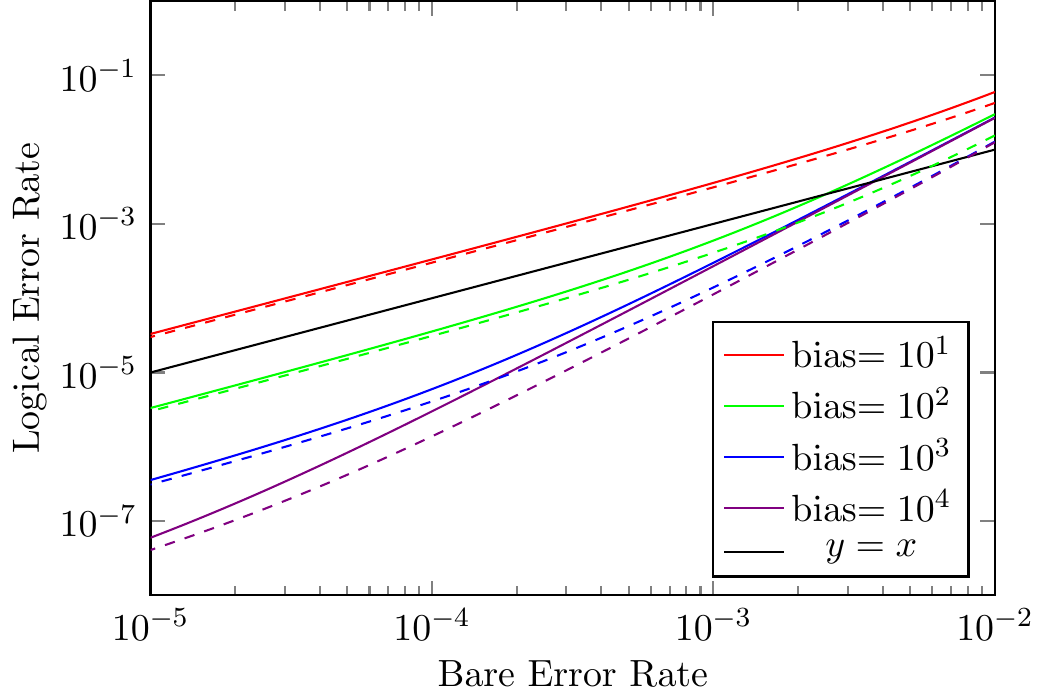}
	\includegraphics[width=0.95\linewidth]{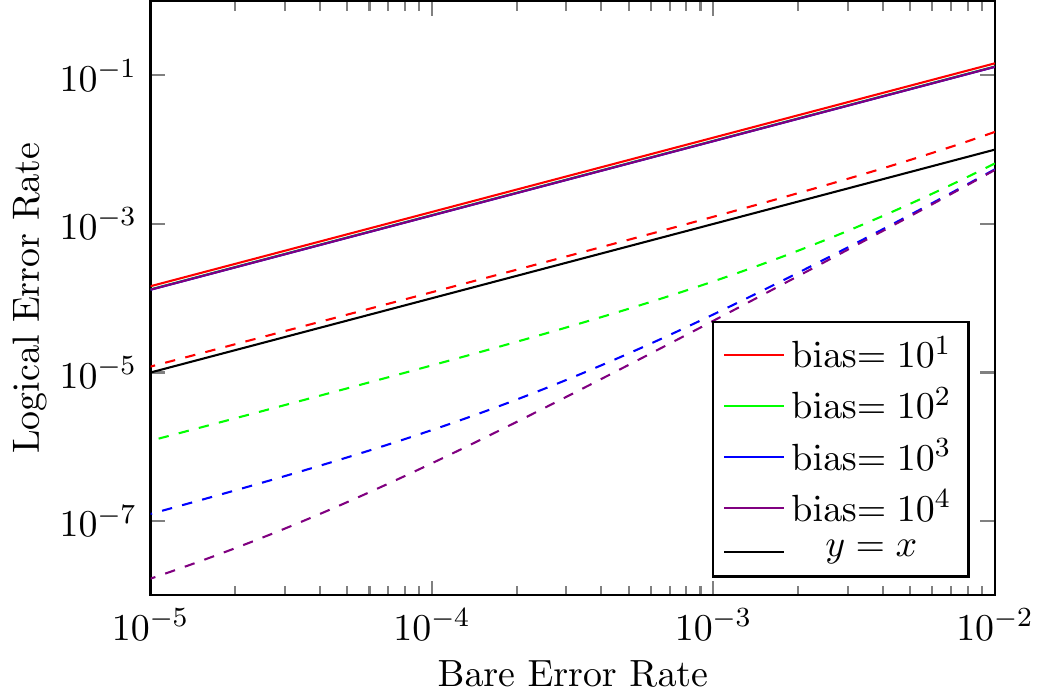}
	\caption{(Color online)  Logical $X_L$ error rates (solid lines) and $Z_L$ error rates (dashed lines) for the gadget with $n=3$ and setting $p_x=p_{zz}$, with number of measurement repetitions $r=n$ (top); with $r=1$ (bottom).  While $r=n$ gives comparable reductions in both $X_L$ and $Z_L$ error rates, the $r=1$ offers more significant reduction in the $Z_L$ error rate while increasing the $X_L$ error rate by about an order of magnitude for all values of bias.  (In the bottom plot, the solid lines are nearly indistinguishable for all values of noise bias.)}
	\label{fig:LogicalNoise}
\end{figure}

The logical $X_L$ and $Z_L$ error rates for various choices of noise bias, for the codes $n=3$ and setting $p_x=p_{zz}$, are shown in Fig.~\ref{fig:LogicalNoise}.  We see that choosing the number of measurement repetitions to be $r=n$, both logical error rates drop below the bare $p_z$ error rate provided that $p_z$ is less than $\sim 3 \times 10^{-3}$ and the bias is greater than $10^2$.  If instead one chooses $r=1$ (no measurement repetitions), the logical $X_L$ error rate increases from the bare $p_z$ error rate by about an order of magnitude, independent of the bias, but the $Z_L$ error rate is reduced by a greater amount than the $r=n$ case.  While at first sight this dependence on $r$ appears to be a nuisance, we will see in the next section how to benefit from this enhanced $Z$ suppression when using our gadget within a distillation protocol.

%The reduction in noise rate achieved by this scheme is illustrated in Fig.~\ref{fig:RepOnlyResults} for the case $n=3$ and choosing $r=n=3$.  We find that for bare error rates of less than $10^{-3}$ and biases of at least $10^2$, the logical error rate of the output magic state is lower than the bare error rate.  That is, in this low error rate regime, high-quality encoded magic states can be prepared using relatively simple circuits, compared with the large overhead associated with magic state distillation schemes.  
%
%
%
%\begin{figure*}
% \centering
%    \includegraphics[width=1.0\textwidth]{RepOnlyResults060415.jpg}
%  \caption{Logical error rate affecting magic states produced encoded in the three qubit repetition code only, as a function of the bare $Z$ error rate for a range of biases. The dotted line is a guide which shows the value of the logical error rate equal to the bare $Z$ error rate. \ptw {Probably want something more formal in legend than ``bias=infinity'' for asymptotic case.}}
%  \label{fig:RepOnlyResults}
%\end{figure*}

\section{Example:  Overhead for distillation}
\label{sec:distill}

The method presented in the previous section provides an approach to preparing encoded magic states with reduced resources in a situation where the noise is biased.  The implication for fault-tolerant quantum computing is that, by being able to prepare encoded magic states with low error rates at the base level of encoding with few resources, the overhead costs at higher levels of encoding are correspondingly reduced.

A full picture of the benefits of this approach for a quantum computing architecture would require detailed simulations.  However, to obtain a lower bound on the potential gains, we can compare our approach to standard techniques for a specific well-studied task such as magic state distillation.  A standard approach in quantifying overhead for distillation is to assume that Clifford gates are ideal and ``free'', and to enumerate the number of non-Clifford gates that are required in the distillation to achieve a desired fidelity.  (We note that such an analysis is somewhat unfair to our scheme, as it ignores the savings in the number of Clifford operations as a result of using this gadget.  We return to this issue at the end of this section.)

With this perspective, we use the gadget for encoded magic states as described in Sec.~\ref{sec:magicprep}, with logical error rates given by Eqns.~(\ref{eq:EXL}) and (\ref{eq:EZL}).  We then perform an analysis based on $|T\rangle$-type magic state distillation using the 15-qubit quantum Reed-Muller (RM) code (see Ref.~\cite{BrooksThesis} for a detailed analysis), concatenated on the repetition code of the gadget presented here.  The ``bare'' error rate for the Reed-Muller code will be the logical error rate from the repetition code.  We make the choice $n=3$ to minimise the overhead, and assume that $p_{zz} = p_x$.

%Performing this a range of bare $Z$ error rates and biases yields the results summarised in Fig.~\ref{fig:RMm}.  These results show that for a bare $Z$ error rate of $10^{-3}$ and a bias of at least $10^3$, the bare error rate can be reduced to approximately $6 \times 10^{-12}$.
%
%\begin{figure*}
% \centering
%    \includegraphics[width=1.0\textwidth]{Results060415.jpg}
%  \caption{Logical error rate affecting magic states produced encoded in the three qubit repetition code, and further encoded using the Reed Muller code, as a function of the bare $Z$ error rate for a range of biases. The dotted line is a guide which shows the value of the logical error rate equal to the bare $Z$ error rate.}
%  \label{fig:RMm}
%\end{figure*}

The overhead for this scheme, quantified as the average number of non-Clifford gates used in the process, can be obtained by counting the number of $CZ(\theta)$ gates (but not CPHASE gates) used on average to prepare an encoded magic state, and then using this number to determine the overhead of the magic state preparation in the quantum RM code.  For $n=3$, the scheme uses 3 $CZ(\theta)$ gates and has a success probability of $3/4$; therefore the average non-Clifford gate cost is $4$.  (The rejection rate is only negligibly increased in the presence of noise.)  The distillation of magic states in the quantum RM code requires 15 copies of the noisy $\ket{T}$ state, and so using the $n=3$ scheme at the physical level and subsequently encoding in an $l$-layer concatenated 15-qubit quantum RM code will have overhead $4 \cdot 15^l$.  (Note that this is ignoring any potential use of non-$|T\rangle$ magic states produced by the gadget when the incorrect $X$ measurement results are obtained.)

%\subsubsection{Balancing error rates}

We have not yet fixed the number of measurement repetitions $r$ in the scheme, and so we now consider the optimal choice for the purposes of RM distillation.  The quantum Reed-Muller code used in this example has a peculiar property from the perspective of biased noise:  it is far more effective at detecting $X$ errors than $Z$ errors.  This imbalance in error correction properties of the quantum RM code means that the logical error rate after concatentation is determined by the gadget's logical $Z_L$ error rate rather than the logical $X_L$ error rate at the lower level.  That is, there is an advantage to sacrificing the logical $X_L$ error rate of Eq.~(\ref{eq:EXL}) to allow for even a relatively modest reduction in the logical $Z_L$ error rate of Eq.~(\ref{eq:EZL}).

\begin{figure}
	\centering
	\includegraphics[width=0.95\linewidth]{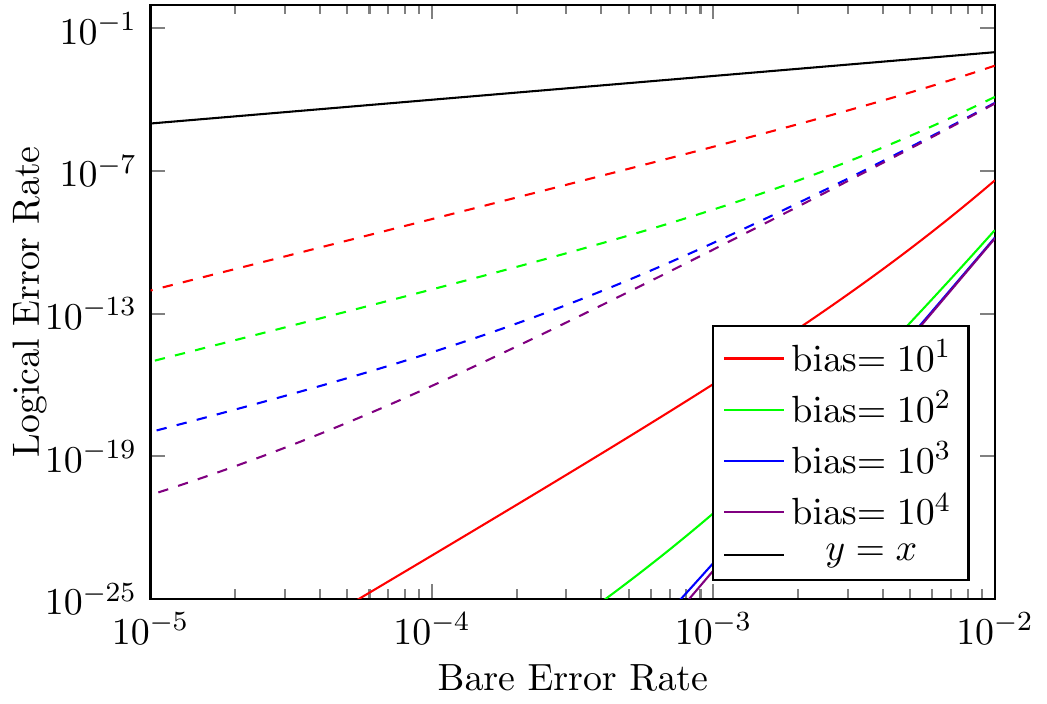}
	\includegraphics[width=0.95\linewidth]{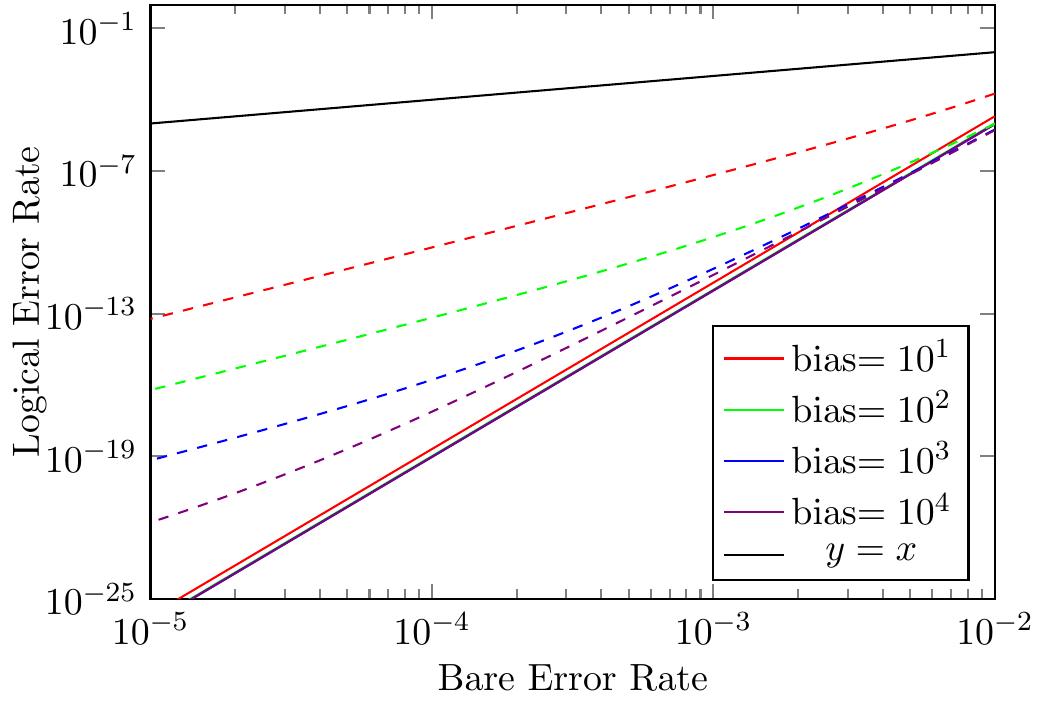}
	\caption{(Color online)  Logical $X_L$ error rates (solid lines) and $Z_L$ error rates (dashed lines) with $n=3$ followed by one layer of concatentation with a 15-qubit quantum Reed-Muller code.  (top) with $r=n$; (bottom) with $r=1$.  }
	\label{fig:LogicalRMNoise}
\end{figure}

Figure~\ref{fig:LogicalRMNoise} presents the final logical error rate of the $n=3$ gadget followed by one layer of concatenation with a 15-qubit quantum RM code.  The choice of number of repetitions $r=1$ (no repetitions) almost always gives a superior reduction in total error rate.  (Note that there is a substantial reduction in the number of Clifford gates by choosing $r=1$ as well, as the optimal choice for noise reduction also minimises the number of measurements, although this additional savings is not captured by this simplistic accounting of overhead using total number of non-Clifford gates.)

Observing the final noise bias affecting the output of the noise-bias gadget concatenated with a 15-qubit quantum RM code with $r=n$ and $r=1$ gives some insight.  As an example, for a bare $Z$ error rate of $10^{-3}$ and bare noise bias of $10^{3}$, choosing $r=n=3$ yields a final noise bias of approximately $4\times10^{12}$, meaning that the logical noise is very strongly dominated by $Z$ errors. By contrast, choosing $r =1$, with the other parameters the same as above, gives an output bias of 0.8, indicating that the final logical $Z$ and $X$ error rates have become comparable.  

To assess the overhead of this noise-bias gadget, we compare with a standard application of the 15-qubit quantum RM code for distillation, where $l'$ layers has an overhead of $15^{l'}$.  Clearly, using the noise-bias gadget will be superior if, for a given target logical error rate, we require $l < l'$, i.e., that the noise-bias gadget at the physical level eliminates the need for at least one layer of concatenation, compared to without the noise-bias gadget.  

\begin{figure}
	\centering
	\includegraphics[width=0.95\linewidth]{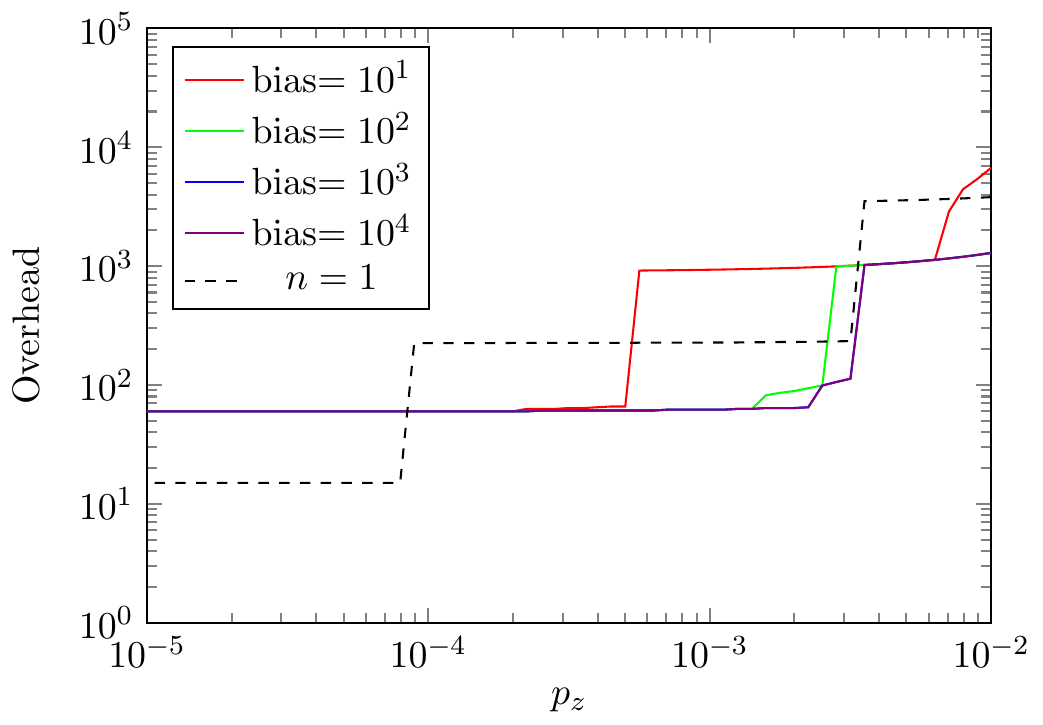}
	\includegraphics[width=0.95\linewidth]{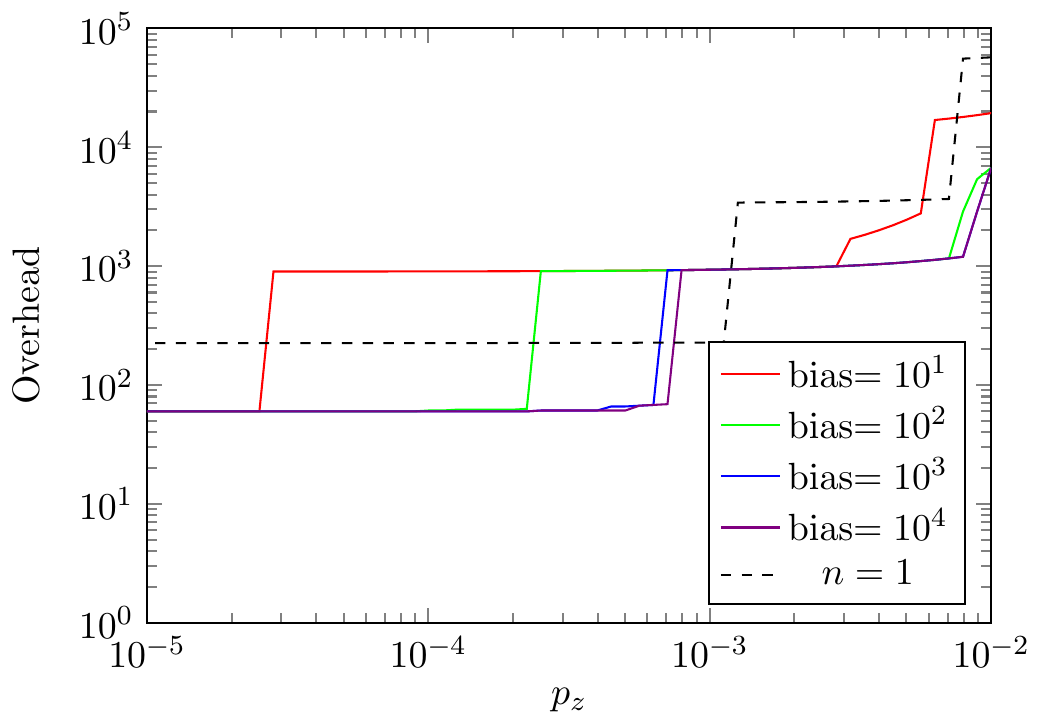}
	\includegraphics[width=0.95\linewidth]{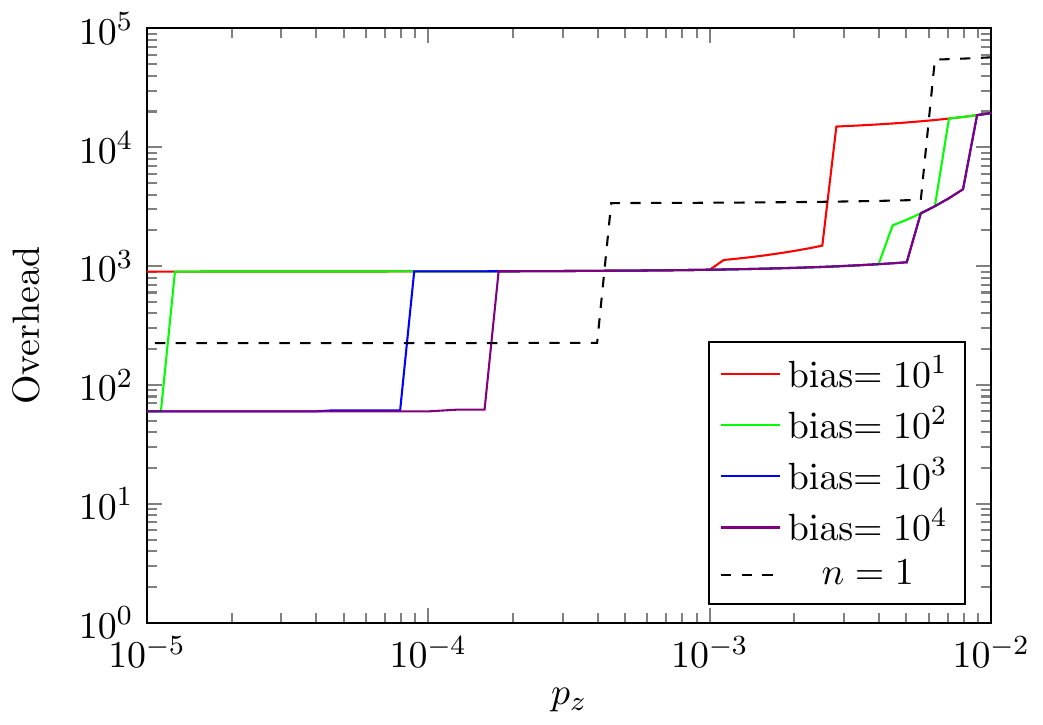}
	\caption{(Color online)  Overhead of the scheme for various values of the bias, for a target error rate of $10^{-8}$ (top), $10^{-12}$ (middle), and $10^{-16}$ (bottom).  For comparison, the overhead for $n=1$ (no noise biased encoding) with the same target error rate is shown.}
	\label{fig:Overheads}
\end{figure}

In Fig.~\ref{fig:Overheads}, we show the overhead required for the noise-bias scheme to achieve a target error rate of $10^{-8}$, $10^{-12}$, and $10^{-18}$ for a variety of noise biases, and compare with the overhead of the standard approach using quantum RM distillation.  The value of $r$ is selected at each point to optimize the scheme, but we note that the choice $r=1$ is nearly always the optimal choice, except for very large bare error rates.  We see that, across a broad range of physical error rates as well as target error rates, using this scheme almost always eliminates the need for one layer of RM code concatenation, with an associated savings in overhead of a factor of $15/4$, provided the bias is greater than 10.  

%Such a trade-off can be accomplished by choosing not to repeat the measurements of $M_Z$ and $M_{ZZ}$, i.e., setting $r = 1$.  Removing these repetitions reduces the number of places $Z$ errors could occur on blocks 1 and 2, at the expense of greatly increasing the probability of a faulty measurement.  This substitution leads to different expressions for the logical error rates of Eqns.~(\ref{eq:EXL}) and (\ref{eq:EZL}), now given by
%\begin{align}
%E_{\bar{X}} &= 5np_x + (3n+4)p_z \,, \\
%E_{\bar{Z}} &= (8p_z)^n + np_{zz} + 3np_x + \sum_{k=1}^{m-1} \binom{n}{k} (4p_z)^{2k} \text{sin}^{2} \left(\frac{k \pi}{4}\right)\,.
%\end{align}
%Again with $n=3$, the performance of the overall scheme is shown as Fig.~\ref{fig:RM1}. These results show that the error rate for, as an example, a bare $Z$ error rate of $10^{-3}$ and bias of at least $10^{3}$ produces a reduction in the logical error rate by a factor of approximately 4.
%\begin{figure*}
% \centering
%    \includegraphics[width=1.0\textwidth]{Results060415requals1.jpg}
%  \caption{Logical error rate affecting magic states produced encoded in the three qubit repetition code, and further encoded using the Reed Muller code, as a function of the bare $Z$ error rate for a range of biases, without repetition of measurements in the $M_Z$ and $M_{ZZ}$ gadgets. The dotted line is a guide which shows the value of the logical error rate equal to the bare $Z$ error rate.}
%  \label{fig:RM1}
%\end{figure*}

Although we have used the number of non-Clifford gates as our measure of `overhead', a more informative measure would be to use the total number of operations including Cliffords.  We can estimate the overheads by this measure as well.  The standard RM distillation scheme uses 139 operations~\cite{BrooksThesis}, whereas the $n=3$ noise-bias gadget uses 41 on average, a savings of a factor of approximately $3.5$.  This savings in overhead is comparable to counting non-Clifford gates:  the standard quantum RM scheme uses 15 non-Cliffords and ours uses $4$ on average, a savings of a factor of $3.75$.

Note also that our method of assessment is unfavorable for the noise-biased scheme, because we account for gate errors in the noise-bias gadget but not in the distillation using the quantum RM code.  For instance, if we compare the use of the noise-bias gadget concatenated with a single quantum RM code with two layers of concatenation of the quantum RM code, then in the former case Clifford errors of the first encoding layer are taken into account while in the latter both layers are assumed to be error-free. For this reason, the actual savings offered by our scheme will be larger.

\section{Conclusion}
\label{sec:conclusion}

We have demonstrated that a method for encoded magic state preparation that operates when the noise is biased can offer a reduction in overhead compared with standard schemes.  The precise gains will depend on the details of the architecture, and in particular the amount of noise and bias at the physical level as well as the desired target logical error rate.  

We briefly consider two regimes of target logical error rate.  First, consider the regime of high noise and a relaxed target of $10^{-8}$, which might be relevant for quantum chemistry calculations or `initial' applications of quantum processors.  In this regime, our approach provides a reduction in overhead by at least a factor of $\sim 4$ (we use our scheme plus one round of RM distillation, as opposed to needing 2 rounds of RM distillation without) for error rates $10^{-4} < p_z < 2 \times 10^{-3}$ provided the bias is greater than $10$.  

Second, consider a more demanding target of say $10^{-16}$, as may be required for large-scale quantum computing.  The noise-bias approach provides a similar improvement in overhead and circuit complexity in the `reasonable' noise regime $4 \times 10^{-4} < p_z < 4 \times 10^{-3}$ for biases greater than $10$, requiring two concatenated quantum RM code layers rather than three.

Our encoded magic-state gadget is explicitly based on the use of a phase-flip repetition code defined by Eq.~(\ref{eq:RepCode}), which offers protection against $Z$ errors but none against $X$ errors.  It would be worthwhile to consider if our results can be generalized to other stabilizer codes that have differing minimum distance for $X$ and $Z$ errors.  Examples of such codes include symmetric Shor codes, Bacon-Shor subsystem codes~\cite{Bacon2006,Brooks}, quantum polar codes, quantum Reed-Muller codes, and Kitaev codes on non-self-dual lattices~\cite{Delfosse}.

\begin{acknowledgments}
The authors are grateful to Andrew Doherty and Steve Flammia for helpful discussions. This work is supported by the ARC via the Centre of Excellence in Engineered Quantum Systems (EQuS) project number CE110001013, and the Intelligence Advanced Research Projects Activity (IARPA) Multi-Qubit Coherent Operations program W911NF-10-1-0330.  The effort depicted is supported in part by the U.S. Army Research Office under contract W911NF-14-C-0048. The content of the information does not necessarily reflect the position or the policy of the Government, and no official endorsement should be inferred.  DP acknowledges the hospitality of the University of Sydney.
\end{acknowledgments}

\end{document}